\documentclass[12pt]{article}

\newcommand{\m}{\mathrm}
\newcommand{\be}{\begin{equation}}
\newcommand{\ee}{\end{equation}}
\newcommand{\ba}{\begin{eqnarray}}
\newcommand{\ea}{\end{eqnarray}}

\usepackage{graphicx}
\usepackage{amssymb}
\usepackage{amsmath}
\usepackage[T1]{fontenc} 
\usepackage[ansinew]{inputenc} 
\usepackage[nosort]{cite}
\newcommand{\inbar}{\vrule height1.57ex width.4pt depth0pt}
\newcommand{\SW}{\relax{\hbox{$\ \inbar\kern-.285em{\rm S}$}}}

\begin{document}
\thispagestyle{empty}
\begin{center}

\null \vskip-1truecm \vskip2truecm

{\Large{\bf \textsf{Cosmic Censorship for AdS$_5$-Kerr}}}

{\large{\bf \textsf{}}}

{\large{\bf \textsf{}}}

\vskip1truecm

{\large \textsf{Brett McInnes}}

\vskip1truecm

\textsf{\\  National
  University of Singapore}

\textsf{email: matmcinn@nus.edu.sg}\\

\end{center}
\vskip1truecm \centerline{\textsf{ABSTRACT}} \baselineskip=15pt
\medskip

We show that cosmic censorship takes an exceptionally complex and interesting form in the case of five-dimensional AdS-Kerr black holes, due to the unusually distant relation obtaining in that case between the black hole parameters and the physical mass and angular momentum. One finds that, in this case, censorship is less restrictive than one might hope: it apparently allows some rather bizarre behaviour, and in particular does not prohibit arbitrarily large angular momenta. We find however that most of the unwelcome geometries permitted by censorship can be eliminated by requiring stability against pair-production of branes. We suggest that the small set of surviving problematic cases can be eliminated in a natural way by imposing a certain (holographic) bound on the physical mass.

\newpage

\addtocounter{section}{1}
\section* {\large{\textsf{1. Censorship for Asymptotically AdS Spacetimes}}}
The question of cosmic censorship, both in four and in higher dimensions, has recently attracted intense interest, in connection with a wide variety of topics of current interest, both theoretical and observational: see \cite{kn:weak,kn:dias,kn:card,kn:rahm,kn:emp,kn:horsant,kn:binwang,kn:freese,kn:zeng} for a selection. Perhaps most striking from a theoretical point of view is the suggestion \cite{kn:weak,kn:horsant} that, in the asymptotically AdS case at least, censorship has a deep relation with the question \cite{kn:palti} as to which effective theories can be lifted to consistent quantum gravity theories in the ultraviolet; that is, that censorship could be one (very easily implemented) component of the selection mechanism which must exist in this context.

Indeed, the case of asymptotically AdS spacetimes takes on a particular aspect, because of the presumed existence (``gauge-gravity duality'' \cite{kn:nat}) of a physically equivalent thermal field theory describing a fluid inhabiting the conformal boundary. If, as has been suggested \cite{kn:nat}, the boundary fluid at all resembles the Quark-Gluon Plasma (studied for example by the STAR collaboration \cite{kn:STAR}), then it equilibrates extremely quickly, so the plasma has, in its generic state, a well-defined temperature and entropy density. In the gauge-gravity picture, this means that the system in the bulk likewise has a well-defined temperature and entropy, which in turn means that it has a well-defined event horizon ---$\,$ think of the familiar Euclidean derivation of the Hawking temperature, which is based on the condition that the geometry should be smooth at the location of the (Euclidean) ``horizon'', and recall that the entropy is proportional to the area of the event horizon. From this point of view, then, a violation of censorship in the asymptotically AdS case would only be possible if the boundary theory can behave in some extremely unexpected way.

However, apparent counter-examples to censorship in the AdS case have been found, culminating in \cite{kn:horsantben,kn:crissant}; but it has been strongly argued \cite{kn:weak,kn:horsant} that these examples are in conflict with the ``weak gravity conjecture'' \cite{kn:palti}. In other words, when considered in a broader context, these alleged counter-examples may not be physical.

Now the results of \cite{kn:weak,kn:horsant} are quite specific: they suggest that, at least in many examples, the weak gravity conjecture yields \emph{precisely} the parameter values needed to preserve censorship. Turning this around, it is natural to suggest that censorship is a necessary, but perhaps not \emph{sufficient}, condition for an effective theory defined on an asymptotically AdS spacetime to avoid being consigned to the ``Swampland'' \cite{kn:swamp}.

This prompts a simple question: what kinds of asymptotically AdS spacetimes \emph{do} satisfy censorship? Are all of them physically acceptable? If not, what exactly prevents them from being so?

Here we will approach this question by studying a very extreme example of an AdS spacetime which nevertheless \emph{does} satisfy censorship. We will focus on five-dimensional black holes, as is natural in the AdS context since the dual theory is then defined in four dimensions. Specifically, we consider AdS$_5$-Kerr spacetimes \cite{kn:mcong,kn:gwak,kn:93}, since these do not involve any matter fields whose behaviour might confuse the issues we consider; that is, they are exactly Einstein manifolds. We will see that, if the angular momentum of such a black hole is very large, the exterior spacetime can behave in a very peculiar way \emph{while being fully compatible with cosmic censorship}. One can argue that this gives explicit expression to the idea that censorship alone does not filter out all candidates for residence in the Swampland.

We begin with a study of the cosmic censorship condition for AdS$_5$-Kerr black holes. It is surprisingly intricate.

\addtocounter{section}{1}
\section* {\large{\textsf{2. Censorship for AdS$_5$-Kerr Black Holes}}}
The AdS$_5$-Kerr metric \cite{kn:hawk,kn:cognola,kn:gibperry}, in the simplest case where one of the two possible rotation parameters, $(a,b)$, is set equal to zero, takes the form
\begin{flalign}\label{A}
g\left(\m{AdSK}_5^{(a,0)}\right)\; = \; &- {\Delta_r \over \rho^2}\left[\,\m{d}t \; - \; {a \over \Xi}\,\m{sin}^2\theta \,\m{d}\phi\right]^2\;+\;{\rho^2 \over \Delta_r}\m{d}r^2\;+\;{\rho^2 \over \Delta_{\theta}}\m{d}\theta^2 \\ \notag \,\,\,\,&+\;{\m{sin}^2\theta \,\Delta_{\theta} \over \rho^2}\left[a\,\m{d}t \; - \;{r^2\,+\,a^2 \over \Xi}\,\m{d}\phi\right]^2 \;+\;r^2\cos^2\theta \,\m{d}\psi^2 ,
\end{flalign}
where
\begin{eqnarray}\label{B}
\rho^2& = & r^2\;+\;a^2\cos^2\theta, \nonumber\\
\Delta_r & = & \left(r^2+a^2\right)\left(1 + {r^2\over L^2}\right) - 2M,\nonumber\\
\Delta_{\theta}& = & 1 - {a^2\over L^2} \, \cos^2\theta, \nonumber\\
\Xi & = & 1 - {a^2\over L^2}.
\end{eqnarray}
Here $L$ is the background AdS curvature length scale, $t$ and $r$ are as usual, and ($\theta, \phi, \psi$) are Hopf coordinates on the topological three-sphere (with $\phi$ and $\psi$ running from $0$ to $2\pi$, while $\theta$ runs from $0$ to $\pi/2$). The quantities $a$ and $M$ (which can be taken non-negative without loss of generality) are to be regarded as strictly geometric parameters (for example, $a$ determines the shape of the event horizon); it is important to understand that, for example, $a$ is \emph{not} interpreted here as the ratio of the angular momentum to the mass of the black hole (as it is in four spacetime dimensions).

Instead (from \cite{kn:gibperry}), we have, if $\mathcal{M}$ denotes the \emph{physical} mass and $\mathcal{J}$ is the angular momentum,
\begin{equation}\label{C}
\mathcal{M}\;=\;{\pi M \left(2 + \Xi\right)\over 4\,\ell_{\textsf{B}}^3\,\Xi^2}, \;\;\;\;\;\mathcal{J}\;=\;{\pi M a\over 2\,\ell_{\textsf{B}}^3\,\Xi^2},
\end{equation}
where $\ell_{\textsf{B}}$ is the gravitational length scale in the bulk. The angular momentum to (physical) mass ratio $\mathcal{A}$ is therefore given by
\begin{equation}\label{D}
\mathcal{A}\;=\;{2 a \over 2 + \Xi}\;=\;{2 a \over 3 - \left(a^2/L^2\right)}.
\end{equation}

Notice that as $a/L$ ranges\footnote{Examining the metric, we see that $a/L$ is not allowed to be \emph{exactly} equal to unity. As we will see later, censorship does not allow $a/L$ even to come close to unity, so we can ignore this technicality here.} in $[\,0,\,\sqrt{3})$, $\mathcal{A}/L$ ranges in $[\,0,\,\infty)$; the right side of (\ref{D}) is a monotonically increasing function, giving a one-to-one correspondence. (Its graph lies below $\mathcal{A} = a$ for $a/L < 1$, above it thereafter; thus $\mathcal{A}/L \,<\,1$ precisely when $a/L \,<\,1$.)

Now we turn to the condition for cosmic censorship to hold for these black holes. The condition is that the equation $\Delta_r = 0$ should have a real, non-negative solution, which we denote by $r_H$. As $\Delta_r$ is a quadratic in $r^2$, it follows (from the observation that a quadratic with all coefficients real and positive cannot have a non-negative root) that censorship for these black holes can be expressed in the deceptively simple form\footnote{Notice that this condition does not depend on $L$, and so it is valid also in the asymptotically flat case. In both situations the extremal case, $a^2 = 2M$, has $r_H = 0$. Requiring as usual that censorship should not be invalidated by small perturbations of the geometry, this is effectively a naked ring singularity \cite{kn:emparanmyers}, not a black hole, and so we exclude this case; that is why we use a strict inequality in (\ref{G}) and subsequently.}
\begin{equation}\label{G}
a^2\;< \;2M;
\end{equation}
This, however, is the formulation in terms of the geometric parameters $a$ and $M$. In terms of the physical parameters $\mathcal{A}$ and $\mathcal{M}$, the condition for censorship is more complicated. For our purposes, it will be useful to choose as our variable $\Xi$ (see the final equation in (\ref{B}) above) to replace $a$. Since $a$ is determined by $\mathcal{A}$, we can think of $\Xi$ as a proxy for $\mathcal{A}$: in detail, we have
\begin{equation}\label{H}
\Xi\;=\;{2L^2\over \mathcal{A}^2}\,\left(\sqrt{1\,+\,{3\mathcal{A}^2\over L^2}}\;-\;1\;-\;{\mathcal{A}^2\over L^2}\right).
\end{equation}
and
\begin{equation}\label{HH}
{\mathcal{A}\over L} \;=\;{2\sqrt{1 - \Xi}\over 2 + \Xi}.
\end{equation}
In terms of this variable, we find (using the first relation in (\ref{C})) that censorship can be stated in terms of physical parameters as\footnote{Here we are of course replacing $a^2$ by $L^2\left(1 - \Xi\right)$, and treating $\Xi$ as a variable representing the variations of $a$, which is only legitimate if $L$ is not infinite (for in that case $\Xi$ has a fixed value, unity, for all values of $a$). Thus, the following discussion does not apply to the asymptotically flat case ($L \rightarrow \infty$), which is why censorship in the AdS case differs so markedly from its counterpart for asymptotically flat Kerr black holes.}
\begin{equation}\label{I}
L^2\left(1 - \Xi\right)\;< \; {8\ell_{\textsf{B}}^3\mathcal{M}\Xi^2\over \pi \left(2 + \Xi\right)},
\end{equation}
which can be re-written as
\begin{equation}\label{J}
\left(\mu + 1\right)\Xi^2\; + \;\Xi \; - 2 \;> \; 0,
\end{equation}
with
\begin{equation}\label{JJ}
\mu \equiv {8\ell_{\textsf{B}}^3\mathcal{M}\over \pi L^2};
\end{equation}
clearly $\mu$ is a kind of dimensionless mass.

The quadratic in (\ref{J}) always has one negative real root, $\Xi_-$, and one positive real root, $\Xi_+$; however we have $\Xi_- \geq - 2$ and $\Xi_+ \leq  1$, with equality in both cases only when $\mu = 0$. Of course, $\Xi_+$ and $\Xi_-$ are simple functions of $\mu$ (only); $\Xi_+$ is decreasing, while $\Xi_-$ is increasing when regarded as such. As a function of $\Xi$, $\mathcal{A}/L$ is strictly decreasing (see (\ref{HH})): it is smaller than unity when $\Xi$ is positive, but greater than unity when $\Xi$ is negative. Therefore we see that censorship requires \emph{either} that $\Xi > \Xi_+$, so that
\begin{equation}\label{JJJ}
{\mathcal{A}\over L} \;<\;{2\sqrt{1 - \Xi_+}\over 2 + \Xi_+} \;<\;1,
\end{equation}
\emph{or}
\begin{equation}\label{JJJJ}
{\mathcal{A}\over L} \;>\;{2\sqrt{1 - \Xi_-}\over 2 + \Xi_-} \;>\;1.
\end{equation}

In short, censorship requires that $\mathcal{A}/L$ be forbidden to come too close to unity, either from below \emph{or from above}. In other words, a band of values around unity is excluded. The right side of (\ref{JJJ}) increases with the dimensionless mass $\mu$, while that of (\ref{JJJJ}) decreases; so the effect of increasing the mass is simply to narrow this band.

To summarize: censorship for these black holes does \emph{not} forbid arbitrarily high angular momenta for a given mass: it only forbids values of $\mathcal{A}/L$ in a band around unity. The mass determines only the \emph{width} of this band: it becomes narrower as the physical mass (and therefore $\mu$) increases. In this, these black holes differ from asymptotically flat five-dimensional Kerr black holes, for which censorship is the simple statement
\begin{equation}\label{M}
\mathcal{A}^2 < {32\ell_{\textsf{B}}^3\mathcal{M}\over 27\pi},
\end{equation}
which does forbid large values of the angular momentum for a given mass\footnote{As is well known \cite{kn:reall}, in six or more dimensions censorship alone does not impose a bound on $\mathcal{A}$ (for fixed $\mathcal{M}$) even in the asymptotically flat case. What we are finding here is that, in the five-dimensional \emph{asymptotically AdS} case, ``large'' values of $\mathcal{A}$ are likewise compatible with censorship. (Here, however, ``large'' means ``large relative to $L$'', not necessarily to $\mathcal{M}$ (or to the horizon radius), so the two situations are not directly comparable. See below for more on this.)}.

To conclude: because of the unusually remote relation between the black hole geometric parameters $M$ and $a$ and the physical parameters $\mathcal{M}$ and $\mathcal{A}$, censorship for AdS$_5$-Kerr black holes takes an unusual and complex form in general. The main point is that censorship appears to impose far weaker restrictions in the asymptotically AdS case than in its asymptotically flat counterpart, in the sense that, for a given mass, only a \emph{finite} range of values for $\mathcal{A}$ is excluded.

The case of $\mathcal{A}/L < 1$ is of course the conventional one; see for example \cite{kn:97} for an application. This is the metric to be used when one imagines obtaining the black hole by gradually ``spinning it up'' from $\mathcal{A} = 0$. However, there is no necessity to confine ourselves to that case: one can consider a black hole which has $\mathcal{A}/L > 1$ ``primordially''.

As we have seen, such a black hole can respect cosmic censorship. Nevertheless, the condition $\mathcal{A}/L < 1$ (or its counterpart in four dimensions) is often imposed; but this does not seem to be a particularly natural restriction. As we know, this is equivalent to requiring that $a/L < 1$. This is usually motivated, perhaps implicitly, by the fact that, if $a/L > 1$, then there is a value of $\theta$ such that the quantity $\Delta_{\theta}$ vanishes at that value; and the reciprocal of $\Delta_{\theta}$ occurs in the AdS$_5$-Kerr metric. Experience with the apparent singularity in the Schwarzschild metric at the event horizon teaches us, however, that this is probably a coordinate problem, and we will show that this is indeed the case. Thus, this is not an argument in favour of $\mathcal{A}/L < 1$.

In fact, one can argue that it is not at all natural simply to impose $\mathcal{A}/L < 1$: we need a clear physical justification. In the gauge-gravity duality, $\mathcal{A}$ corresponds to the ratio of the angular momentum density to the energy density of the fluid represented by the boundary field theory. However, in that theory, $L$ has an interpretation in terms of $N_c$, the number of colours: one has $\ell_{\textsf{B}}^3/ L^3 = \pi/2N_c^2$ \cite{kn:nat}. In the holographic picture, then, insisting on $\mathcal{A}/L < 1$ amounts to postulating that the specific angular momentum of the fluid is bounded by a multiple of $\ell_{\textsf{B}}N^{2/3}_c$. That would be remarkable; if it exists, such a bound is clearly something that needs to be demonstrated and clearly understood, not simply imposed because it is convenient.

Again, if we turn to the Euclidean version of the AdS$_5$-Kerr geometry, we see at once from both of the terms in square brackets in (\ref{A}) that, if $t$ is to be complexified, then so also must $a$ be: but then the Euclidean version of $\Delta_{\theta}$, $\Delta^E_{\theta}$, is given by
\begin{equation}\label{MM}
\Delta^E_{\theta}\; =  \;1\, +\, {a^2\over L^2} \, \cos^2\theta,
\end{equation}
which is obviously positive for all $\theta$, whatever the value of $a/L$. Thus, from the Euclidean point of view, requiring $a/L$ or $\mathcal{A}/L$ to be smaller than unity is not well-motivated.

We conclude, then, that some concrete physical argument is needed to justify the exclusion of black holes with $\mathcal{A}/L > 1$. Let us therefore consider this case more carefully.

\addtocounter{section}{1}
\section* {\large{\textsf{3. AdS$_5$-Kerr Black Holes with $\mathcal{A}/L > 1$}}}
The fact that AdS$_5$-Kerr spacetimes can satisfy cosmic censorship even when $\mathcal{A}/L$ is large makes that case very interesting, but it does not prove that such objects can really be physical. In this section we will consider this question from several different points of view.

[1] As mentioned above, the metric is apparently singular at a particular value of $\theta$. We will see that this is an illusion, but we still need to investigate the strange region demarcated by that value.

[2] Intuition suggests that large values of $\mathcal{A}/L$ (and $a/L$) could mean that the black hole is rotating so rapidly that it becomes unstable. This could happen in several different ways. For example, it is well known that, in the asymptotically flat case in dimensions higher than five, ``ultra-spinning'' black holes become unstable because the event horizon becomes so flattened that the black hole resembles a black brane, and these have well-studied classical instabilities \cite{kn:reall}. Notice however that we are interested in the case where $a$ is large relative to $L$, not to the event horizon parameter $r_H$. It is possible to construct AdS$_5$-Kerr black holes such that $a$ is large relative to $L$, yet comparable to $r_H$ (one has only to choose $M$ appropriately). Thus, this effect need not arise here. A more interesting question is to ask what happens when waves scatter off the black hole: this can induce an instability, known as \emph{superradiance}, and it can occur in the AdS-Kerr context \cite{kn:hawkreall}; so we need to consider this.

[3] Even if the AdS$_5$-Kerr spacetimes with large $\mathcal{A}/L$ are acceptable at the perturbative level, one must bear in mind that the full theory in the bulk is not pure Einstein gravity: it is (some limiting case of) string theory. It is well known \cite{kn:84} that non-perturbative string effects can destabilize apparently innocuous asymptotically AdS black hole spacetimes, and so it is essential to ask whether this is relevant here.

We consider these questions in turn.

\subsubsection*{{\textsf{3.1 The Spacetime Geometry when $\mathcal{A}/L > 1$ }}}
When $\mathcal{A}/L > 1$, we have also $a/L > 1$. This means that $\Xi < 0$; it also means that $\Delta_{\theta}$ vanishes at $\theta = \theta_a$, given by
\begin{equation}\label{MMM}
\theta_a\;\equiv \;\arccos\left(L/a\right).
\end{equation}
For values of $\theta$ smaller than this, $\Delta_{\theta}$ is negative.

From a holographic point of view, the directly physical four-dimensional spacetime is (approximated by) the conformal boundary, so let us focus on the possibility of unphysical behaviour there.

From (\ref{A}) we have at infinity
\begin{flalign}\label{N}
g\left(\m{AdSK}_5^{(a,0)}\right)_{\infty} \;=\; - \left[\,\m{d}t \; - \; {a \over \Xi}\,\sin^2\theta \,\m{d}\phi\right]^2\;+\;{L^2 \m{d}\theta^2  \over \Delta_{\theta}}\;+\;{L^2\sin^2\theta \Delta_{\theta}\,\m{d}\phi^2\over \Xi^2} \;+\;L^2\cos^2\theta \,\m{d}\psi^2 ,
\end{flalign}
where the conformal factor has been chosen so that $t$ represents proper time for a distinguished observer.

We see that there is apparent misbehaviour at $\theta = \theta_a$, where $\Delta_{\theta}$ vanishes. As might be expected, however, this is not a real singularity; no curvature invariants diverge there. For example, the square of the curvature tensor (Kretschmann invariant) is given by
\begin{equation}\label{O}
R^{abcd}R_{abcd}\;=\;{20a^2\over L^6}\,\cos^2\theta\,\left[{3a^2\over L^2}\cos^2\theta\,-\,2\,-\,{2a^2\over L^2}\right]\;+\;{4\over L^4}\,\left[3\,+\,{2a^2\over L^2}\,+\,{3a^4\over L^4}\right],
\end{equation}
which evidently does not diverge anywhere. The same is true of all other curvature invariants: $\theta = \theta_a$ is no more singular than the event horizon; there is no problem here.

Inside $\theta = \theta_a$, $\Delta_{\theta}$ is negative. This means that the second and third terms in the metric, like the first, become negative-definite: the signature of the metric becomes $\left(-,\,-,\,-,\,+\right)$ instead of $\left(-,\,+,\,+,\,+\right)$, that is, it remains Lorentzian, just with a different convention. In other words, $t$ has become spacelike, $\phi$ and $\theta$ remain spacelike, but $\psi$ is now timelike.

Thus, the region $\theta < \theta_a$ is in this way similar to the interior of an event horizon. However, it differs in two important ways. First, of course, it contains no true singularity. Second, it does not forbid signals from escaping. For example, consider the curve, with parameter $s$, given by
\begin{eqnarray}\label{P}
t& = & 0, \nonumber\\
\phi & = & 0,\nonumber\\
\psi& = & s, \nonumber\\
{\sin \theta\over \sqrt{{a^2\over L^2}\,\cos^2 \theta - 1}}\;-\;{\sqrt{{a^2\over L^2}\,\cos^2 \theta - 1}\over \sin \theta} & = & 2\,\tan\left(2s\right).
\end{eqnarray}
It is not difficult to show that this is a null geodesic, extending from $\theta = 0$ (where $\psi = -\pi/4$) to $\theta = \theta_a$ (where $\psi = \pi/4)$. Thus a ray of light can reach $\theta = \theta_a$ in a proper time, according to an observer at $\theta = 0$, which is finite (equal to $\pi L/2$, one quarter of the circumference of the circle there parametrised by $\psi$). From there it can continue on into the outside world: $\theta = \theta_a$ is \emph{not} an event horizon.

Since $\psi$ is an angular coordinate (ranging, like $\phi$, from zero to $2\pi$), we see that the region $\theta < \theta_a$ contains closed timelike worldlines. One might suggest trying to avoid this by ``unwinding'' the periodic timelike coordinate, as indeed one does with AdS spacetime itself; but that cannot be done here, since we need the $\theta < \theta_a$ region to attach continuously to the remainder of the spacetime, in which $\psi$ certainly is periodic. In short, this region of the spacetime does contain closed timelike worldlines; and, as we have just seen, the corresponding acausal behaviour is fully visible from outside.

The status of spacetimes with closed timelike worldlines remains a vexed question. Some authors \cite{kn:kip} are tentatively willing to accept them, while others \cite{kn:matt} take the view that they should be excluded, \emph{but} only if one can find a clear physical justification for doing so. The argument is that, while such spacetimes may exist as solutions in classical General Relativity, they are not valid solutions in quantum gravity: they belong to the Swampland. In this work, we will follow this approach; that is, rather than reject such spacetimes out of hand, we will search for concrete reasons to argue that \emph{they do not exist as valid solutions in string theory}. In the meantime, however, the point we wish to stress here is that, in the AdS$_5$-Kerr context, cosmic censorship alone does \emph{not} rule out such behaviour.

Before proceeding to that, let us see whether these solutions can be ruled out in a simpler way: perhaps the black hole ``rotates too quickly to be stable''.

\subsubsection*{{\textsf{3.2 Superradiance }}}
For a particle with angular momentum per unit mass $\mathcal{H}$ (relative to this coordinate system) on an equatorial orbit, outside or on the event horizon of an AdS$_5$-Kerr black hole, we have (from equation (\ref{A}))
\begin{equation}\label{Q}
{a\over r^2 \Xi} \left(\Delta_r - \left[r^2 + a^2\right]\right)\dot{t}\;+\;{1\over r^2 \Xi^2}\left(-\, a^2 \Delta_r + \left[r^2 + a^2\right]^2\right)\dot{\phi}\;=\;\mathcal{H}.
\end{equation}
In particular, for a (necessarily massless) particle with zero angular momentum, which remains on the event horizon, we have
\begin{equation}\label{QQ}
{\m{d}\phi\over \m{d}t}\;=\;{a\,\Xi\over r_H^2 + a^2}.
\end{equation}
Relative to a particle at infinity which also has zero angular momentum (such a particle has angular velocity $- a/L^2$ in these coordinates), such a particle has angular velocity
\begin{equation}\label{QQQ}
\omega\;=\;{a\,\left(1 + {r_H^2\over L^2}\right)\over r_H^2 + a^2}:
\end{equation}
this is sometimes called the ``angular velocity of the horizon''.

It is known \cite{kn:hawkreall} that these black holes are stable against a superradiant instability if $\omega L < 1$, and it is conjectured that this condition is also necessary. Assuming that this is so, the condition for stability in this case is
\begin{equation}\label{R}
{{a\over L}\,\left(1 + {r_H^2\over L^2}\right)\over {r_H^2\over L^2} + {a^2\over L^2}}\;<\;1.
\end{equation}
Intuitively, one might have thought that the superradiant instability would be exacerbated by large values of $\mathcal{A}/L$ or $a/L$, but we see now that this is not the case: the expression on the left in (\ref{R}) is actually \emph{small} when $a/L$ is large. However, it turns out that this condition can indeed be violated for moderate values of $a/L >1$; on the other hand, it can also be violated by values of $a/L < 1$.

To take an artificial but instructive example: suppose we fix $r_H^2/L^2 = 0.5$. (This can be done by systematically choosing $M$ to enforce it, in the equation for $r_H$; notice that this procedure automatically ensures that censorship is satisfied.) Then one actually finds that (\ref{R}) is satisfied for \emph{all} $a/L > 1$, but it is \emph{violated} for all $a/L$ between 0.5 and unity. Other examples, however, have the reverse properties.

The perhaps counter-intuitive conclusion is that, while the superradiant instability can indeed arise when $\mathcal{A}/L$ is larger than unity, it need not: these objects can be classically stable, though of course one must verify this in each individual case.

It remains to consider non-classical sources of instability. Since, in the AdS/CFT duality, the bulk physics is ``stringy'', the natural way to approach this is to consider restrictions arising from instabilities of that kind. One particularly interesting and powerful such restriction is the one considered by Seiberg and Witten in \cite{kn:seiberg}. We will now show that this criterion does in fact severely restrict the possible physical values of $\mathcal{A}/L$.

\subsubsection*{{\textsf{3.3 The Seiberg-Witten Stability Criterion }}}
Seiberg and Witten \cite{kn:seiberg} (see also \cite{kn:maldacena,kn:KPR}) studied a fundamental question in string-theoretic physics on ($d+1$)-dimensional asymptotically AdS spacetimes: is the system stable against the nucleation of \emph{branes}? They showed that the situation is particularly delicate for BPS branes. In this case, stability requires that the following condition be satisfied. (We follow Seiberg and Witten and work in Euclidean signature, converting to Lorentzian signature at the end of the computation. In fact the Euclidean version of the condition we study has a deep interpretation in terms of the internal mathematical consistency of string theory on asymptotically hyperbolic manifolds: see \cite{kn:ferrari1,kn:ferrari2,kn:ferrari3,kn:ferrari4}.)

Consider any hypersurface $\Sigma$ in the bulk homologous to the conformal boundary, with area $A(\Sigma)$ containing a volume $V(M_{\Sigma})$. Then the condition that the system be stable against nucleation of branes is simply
\begin{equation}\label{S}
A(\Sigma)\;- \;{d \over L}V(M_{\Sigma}) \;\geqslant \;0,
\end{equation}
This condition is certainly satisfied by all surfaces in Euclidean $AdS_{d+1}$ homologous to the boundary, and similarly for the Euclidean AdS-Schwarzschild geometry when the event horizon has a spherical topology.

When converted back to Lorentzian signature, the quantity on the left in (\ref{S}) is just a constant multiple of the action of an arbitrarily located BPS brane. The condition is then simply that the action should not be negative: particularly, it should not be unbounded below.

Let us consider this condition in the present case. The Euclidean version of the AdS$_5$-Kerr metric is given by
\begin{flalign}\label{T}
g^E\left(\m{AdSK}_5^{(a,0)}\right)\; = \; & {\Delta^E_r \over \rho_E^2}\left[\,\m{d}t \; + \; {a \over \Xi^E}\,\m{sin}^2\theta \,\m{d}\phi\right]^2\;+\;{\rho_E^2 \over \Delta^E_r}\m{d}r^2\;+\;{\rho_E^2 \over \Delta^E_{\theta}}\m{d}\theta^2 \\ \notag \,\,\,\,&+\;{\m{sin}^2\theta \,\Delta^E_{\theta} \over \rho_E^2}\left[a\,\m{d}t \; - \;{r^2\,-\,a^2 \over \Xi^E}\,\m{d}\phi\right]^2 \;+\;r^2\cos^2\theta \,\m{d}\psi^2 ,
\end{flalign}
where, here and henceforth, the superscript (or subscript) ``E'' denotes a Euclidean quantity, and where
\begin{eqnarray}\label{U}
\rho_E^2& = & r^2\;-\;a^2\cos^2\theta, \nonumber\\
\Delta^E_r & = & \left(r^2-a^2\right)\left(1 + {r^2\over L^2}\right) - 2M,\nonumber\\
\Delta^E_{\theta}& = & 1 + {a^2\over L^2} \, \cos^2\theta, \nonumber\\
\Xi^E & = & 1 + {a^2\over L^2}.
\end{eqnarray}
As usual, the event horizon is replaced in the Euclidean section by an ``origin of coordinates'', $r_O$, given by solving $\Delta^E_r(r_O) = 0$; then $r \geq r_O \geq a$. (This ensures that $\rho_E$ is real.)

In this geometry, the area of a surface defined by a specific value of $r$ is
\begin{equation}\label{V}
A(r)\;=\;{K\,r\,\sqrt{\Delta^E_r}\over \Xi^E}\int_0^{\pi/2}\rho_E \,\sin \theta\, \cos \theta\, \m{d}\theta,
\end{equation}
while the volume enclosed is
\begin{equation}\label{W}
V(r)\;=\;{K\over \Xi^E}\int_{r_O}^r\int_0^{\pi/2}r\,\rho_E^2\,\sin \theta\, \cos \theta \,\m{d}\theta \,\m{d}r;
\end{equation}
here $K$ is a positive constant determined by the ($r$-independent) periods of $t$ (which is compactified, as usual in Euclidean gravity), $\phi$, and $\psi$; its value is not important here.

The integrals are straightforward. Performing them, substituting into (\ref{S}), and converting back to Lorentzian signature, we obtain a Lorentzian brane action
\begin{equation}\label{X}
\mathfrak{S}\;\propto\;{r\,\sqrt{\Delta_r}\over 3 a^2}\left(r^3 - \left[r^2 - a^2\right]^{3/2}\right)\;-\;{1\over 2L}\left(r^4 + r^2a^2 - r^4_H - r^2_Ha^2\right),
\end{equation}
where the constant of proportionality is positive.

For our purposes, it suffices to understand this function at large values of $r$.

Because of the peculiarities of hyperbolic geometry, both the area and the volume grow, at large $r$, at a rate given asymptotically by a multiple of $r^4$. With the coefficients as in equation (\ref{S}), however, these dominant terms cancel (as Seiberg and Witten emphasise, this behaviour is generic, see \cite{kn:seiberg}), and so the behaviour of the action function at large $r$ is determined by the quadratic term; one finds that
\begin{equation}\label{Y}
\mathfrak{S}\;\propto\;\left(1 - {a^2\over 2L^2}\right)r^2 + C + {D\over r^2} + \cdots,
\end{equation}
where $C$ and $D$ are constants which depend on $a$, $L$, and also $M$; the ellipsis denotes terms which are negligible at large $r$; again, the constant of proportionality is positive.

We see that, if $a^2/L^2$ exceeds 2, then the action is unbounded below\footnote{If $a^2/L^2$ is exactly equal to 2, then the asymptotically dominant term is the constant one. Thus the action is not unbounded below. On the other hand, the constant can be negative; this depends on the relation between the parameters $a$, $L$, and also $M$, which does appear at this level. (The event horizon radius also appears, but it is determined by these quantities.) This case is therefore not clear-cut, but for definiteness we allow it. See \cite{kn:maldacena} for a discussion of a similar situation.} at large $r$, and there will be uncontrolled pair-production of branes (see \cite{kn:maldacena}). \emph{These spacetimes are therefore unacceptable in the full AdS/CFT correspondence.}

We know that the range of $a/L$ (leaving censorship to one side for a moment) is $[\,0,\,\sqrt{3})$, so the restriction to $[\,0,\,\sqrt{2}]$ may not seem to be very substantial. In terms of the physical parameter $\mathcal{A}$, however, this is in fact a very drastic restriction: instead of ranging in $[\,0,\,\infty)$, $\mathcal{A}/L$ can now only range in the finite, and indeed \emph{short}, interval $[\,0,\,2\sqrt{2}]$.

This restriction is strong in another sense: it means that $\mathcal{A}$ is bounded by $2\sqrt{2}L$, and in a holographic picture this kind of restriction, as we pointed out earlier, imposes an interesting relation between $\mathcal{A}$ and the number of colours in the holographically dual theory on the boundary. Specifically, since $\ell_{\textsf{B}}^3/ L^3 = \pi/2N_c^2$ in the holographic dictionary \cite{kn:nat}, Seiberg-Witten instability in this case can be avoided only if $\mathcal{A}$ satisfies
\begin{equation}\label{YY}
\mathcal{A}\;\leq\;2\sqrt{2}\left({2N_c\over \pi}\right)^{1/3}\ell_{\textsf{B}}.
\end{equation}
This throws a remarkable new light on the familiar assertion \cite{kn:nat} that, for the gauge-gravity duality to be useful, $N_c$ must be large; for one certainly expects $\mathcal{A}$ to be capable of reaching values many orders of magnitude larger than $\ell_{\textsf{B}}$, the bulk gravity length scale. In any case, we see that, within string theory, the specific angular momentum of such a black hole is indeed bounded above ---$\,$ though not by censorship. (Compare (\ref{YY}) with the bound (\ref{M}) in the asymptotically flat case.)

In the case where $\mathcal{A}/L > 1$, the range for $\mathcal{A}/L$ is of course narrower still; and if we require that censorship should hold, it narrows even further. Combining our results here with (\ref{JJJJ}), we have now
\begin{equation}\label{Z}
1\;<\;{2\sqrt{1 - \Xi_-}\over 2 + \Xi_-}\;<\;{\mathcal{A}\over L} \;\leq\; 2\sqrt{2},
\end{equation}
where we remind the reader that $\Xi_-$ is the negative root of the quadratic in (\ref{J}). This is a narrow range indeed for most values of the parameters.

Nevertheless, it seems that we have not quite excluded black holes with $\mathcal{A}/L > 1$: in some cases, there is a narrow window of parameter values such that the black hole seems to be able to evade both superradiant and Seiberg-Witten instabilities. This does not seem reasonable; one naturally asks: can the window be closed altogether?

\subsubsection*{{\textsf{3.4 A Conjectural Bound on the Physical Mass }}}
In fact, there is a simple and natural way of achieving this. For, in some cases, the interval in (\ref{Z}) shrinks to \emph{zero}: clearly the lower bound meets the upper bound when $\Xi_- = -1$, which means simply that the dimensionless mass parameter $\mu$ introduced earlier (equation (\ref{JJ})) satisfies $\mu = 2$. Since the lower bound rises as $\mu$ \emph{decreases}, this means that our exotic solutions with $\mathcal{A}/L > 1$ \emph{are ruled out} if $\mu \leq 2$.

Now ruling out these bizarre objects is, we would argue, highly desirable; and we now have the means to do so. That is, we conjecture that a more elaborate analysis of the string-theoretic constraints on the bulk physics will \emph{require}
\begin{equation}\label{ALPHA}
\mu \;\leq \;2.
\end{equation}
In terms of the physical mass of the black hole, this is (from (\ref{JJ}))
\begin{equation}\label{BETA}
\mathcal{M} \;\leq\; {\pi L^2\over 4 \ell_{\textsf{B}}^3};
\end{equation}
or, using the holographic dictionary once again,
\begin{equation}\label{GAMMA}
\mathcal{M} \;\leq\; \left({\pi \over 16}\right)^{1/3}{N_c^{4/3}\over \ell_{\textsf{B}}}.
\end{equation}
This is not an unreasonable condition, in the sense that one does expect the right side to be large, and so $\mathcal{M}$ is not unduly restricted. On the other hand, it is clear that this condition does not involve the angular momentum, so it would have to be imposed on \emph{all} asymptotically AdS$_5$ black holes. It remains to be seen whether the condition is reasonable from this point of view.

Notice that the condition (\ref{ALPHA}) has important consequences for the possible values of $\mathcal{A}/L$ when it is \emph{smaller} than unity. For we recall that, in that case, cosmic censorship requires that $\mathcal{A}/L$ should be bounded above by an expression (see (\ref{JJJ})) involving $\Xi_+$, the positive root of the quadratic in (\ref{J}). When $\mu = 2$, one finds that $\Xi_+ = 2/3$; inserting this into (\ref{JJJ}), and recalling that this upper bound decreases as $\mu$ becomes smaller, we have now
\begin{equation}\label{DELTA}
{\mathcal{A}\over L}\;<\;{\sqrt{3}\over 4}\;\approx\;0.433.
\end{equation}
If we accept this, then we can strengthen the bound on $\mathcal{A}$ given above, (\ref{YY}), very considerably:
\begin{equation}\label{EPSILON}
\mathcal{A}\;\leq\;{\sqrt{3}\over 4}\left({2N_c\over \pi}\right)^{1/3}\ell_{\textsf{B}}.
\end{equation}
Again, the point is that $\mathcal{A}$ is in fact bounded for these black holes, once they are embedded in string theory.
\section* {\large{\textsf{4. Conclusion}}}
There are strong indications that cosmic censorship plays a key role in understanding a crucial question: which effective theories arise from consistent quantum gravity theories in the ultraviolet? Here we have seen an important example, rotating asymptotically AdS black holes in five bulk dimensions; for these, censorship alone is surprisingly weak. It can however be supplemented by invoking non-perturbative effects in the bulk. Arguing in this way has led us to interesting ``holographic bounds'' on the specific angular momentum and even, conjecturally, on the physical masses of all asymptotically AdS$_5$ black holes. If this latter conjecture is wrong, one will need to consider how else one might eliminate the peculiar black holes which satisfy the conditions specified in (\ref{Z}).

\addtocounter{section}{1}
\section*{\large{\textsf{Acknowledgement}}}
The author is grateful to Dr. Soon Wanmei for useful discussions.

\end{document}